\documentclass[a4paper,10pt,twoside]{cpc-hepnp}

\usepackage{multicol}
\usepackage{graphicx}
\usepackage{booktabs}
\usepackage{amssymb,bm,mathrsfs,bbm,amscd}
\usepackage[tbtags]{amsmath}
\usepackage{lastpage}

\begin{document}

\fancyhead[c]{\small Chinese Physics C~~~Vol. XX, No. X (201X)
XXXXXX} \fancyfoot[C]{\small 010201-\thepage}

\footnotetext[0]{}

\title{Proton irradiation effect on SCDs\thanks{Partially supported by National Natural Science
Foundation of China (10978002) }}

\author{%
      YANG Yan-Ji(ÑîÑåÙ¥)$^{1,2;1)}$\email{yangyj@mail.ihep.ac.cn}%
\quad LU Jing-Bin(½¾°±ò)$^{1}$
\quad WANG Yu-Sa(ÍõÓÚØí)$^{2}$
\quad CHEN Yong(³ÂÓÂ)$^{2}$\\
\quad XU Yu-Peng(ÐìÓñÅó)$^{2}$
\quad CUI Wei-Wei(´Þέέ)$^{2}$
\quad LI Wei(Àîì¿)$^{2}$
\quad LI Zheng-Wei(ÀîÕýΰ)$^{2}$\\
\quad LI Mao-Shun(Àîï˳)$^{2}$
\quad LIU Xiao-Yan(ÁõÏþÑÞ)$^{1,2}$
\quad WANG Juan(Íõ¾ê)$^{2}$
\quad HAN Da-Wei(º«´óì¿)$^{2}$\\
\quad CHEN Tian-Xiang(³ÂÌïÏé)$^{2}$
\quad LI Cheng-Kui(Àî³Ð¿ü)$^{2}$
\quad HUO Jia(»ô¼Î)$^{2}$
\quad HU Wei(ºúμ)$^{2}$\\
\quad ZHANG Yi(ÕÅÒÕ)$^{2}$
\quad LU Bo(½²¨)$^{2}$
\quad ZHU Yue(Öì«h)$^{2}$
\quad MA Ke-Yan(Âí¿ËÑÒ)$^{1}$
\quad Wu Di(ÎâµÛ)$^{2}$\\
\quad LIU Yan(Áõçü)$^{2,3}$
\quad ZHANG Zi-Liang(ÕÅ×ÓÁ¼)$^{2}$
\quad YIN Guo-He(Òü¹úºÍ)$^{2}$
\quad Wang Yu(ÍõÓî)$^{2}$
}
\maketitle

\address{%
$^1$ College of Physics, Jilin University, No.2699, Qianjin Road, Changchun 130023, China\\
$^2$ Key Laboratory for Particle Astrophysics, Institute of High Energy Physics, Chinese Academy of Sciences (CAS), 19B Yuquan
Road, Beijing 100049, China\\
$^3$ School of Physical Science and Technology,Yunnan University, Cuihu North Road 2, Kunming, 650091, China\\
}

\begin{abstract}
The Low Energy X-ray Telescope is a main payload on the Hard X-ray Modulation Telescope satellite. The swept charge device is selected for the Low Energy X-ray Telescope. As swept charge devices are sensitive to proton irradiation, irradiation test was carried out on the HI-13 accelerator at the China Institute of Atomic Energy. The beam energy was measured to be 10 MeV at the SCD. The proton fluence delivered to the SCD was $3\times10^{8}\mathrm{protons}/\mathrm{cm}^{2}$ over two hours. It is concluded that the proton irradiation affects both the dark current and the charge transfer inefficiency of the SCD through comparing the performance both before and after the irradiation. The energy resolution of the proton-irradiated SCD is 212 eV@5.9 keV at $-60\,^{\circ}\mathrm{C}$, while it before irradiated is 134 eV. Moreover, better performance can be reached by lowering the operating temperature of the SCD on orbit.
\end{abstract}

\begin{keyword}
SCD, HXMT, proton irradiation, energy resolution, readout noise
\end{keyword}

\begin{pacs}
29.40.Wk.
\end{pacs}

\footnotetext[0]{\hspace*{-3mm}\raisebox{0.3ex}{$\scriptstyle\copyright$}2013
Chinese Physical Society and the Institute of High Energy Physics
of the Chinese Academy of Sciences and the Institute
of Modern Physics of the Chinese Academy of Sciences and IOP Publishing Ltd}%

\begin{multicols}{2}

\section{Introduction}

The Low Energy X-ray Telescope (LE) is a main payload on the Hard X-ray Modulation Telescope (HXMT) satellite\cite{lab1}, which is planned to be launched in about 2015 with a
planned mission duration of 4 years in a 550 km circular orbit around the earth. The LE is required to detect the X-rays from 1 to 15 keV with the energy resolution of the full width at the half maximum (FWHM) better than 450 eV@5.9 keV and the time resolution no more than 1ms. In that way, the SCD \cite{lab2} (Swept Charge Devices) is selected for LE.

SCD is a new type of charge coupled device (CCD). The dark current of the SCD is much lower by being operated in the inverted mode operation (IMO)\cite{lab3}. The SCD gains the faster readout speed by abandoning the position information, and the readout time of the whole frame can be less than 1ms\cite{lab3}. The SCD (CCD236) is especially designed to meet the scientific objectives of LE. It is also used on Chandrayaan-2\cite{lab6}.

In space, there are many protons of all energy, which can cause the decline of the energy resolution of the CCDs. Radiation received by the CCD is specified by two different units. For bulk damage effects, the incoming radiation is specified as a fluence, which increases the charge transfer inefficiency (CTI). For ionization damage, radiation is specified in terms of the total dose, which increases the dark current\cite{lab7}. For LE, the total proton dose received by the detector in 4 years will be 10 MeV-equivalent fluence of $3\times10^{8}\mathrm{p}/\mathrm{cm}^{2}$. To evaluate the performance of the SCD after irradiating the total dose of the four-year mission, a proton irradiation experiment has been conducted on the HI-13 accelerator at the China Institute of Atomic Energy.

\section{Experimental arrangement}

The proton irradiation experiment was carried out on the HI-13 accelerator at the China Institute of Atomic Energy in May, 2012. The SCD was fixed at the back of a plastic plate, which was installed on the mounting plate. The SCD was linked to the pre-amp board with a flexible board inside the chamber. The readout system outside the chamber supplied power to the SCD via sub-D connectors and received the signals from the SCD by Bayonet Nut Connectors (BNC).
\begin{center}
\includegraphics[origin=c,width=8cm]{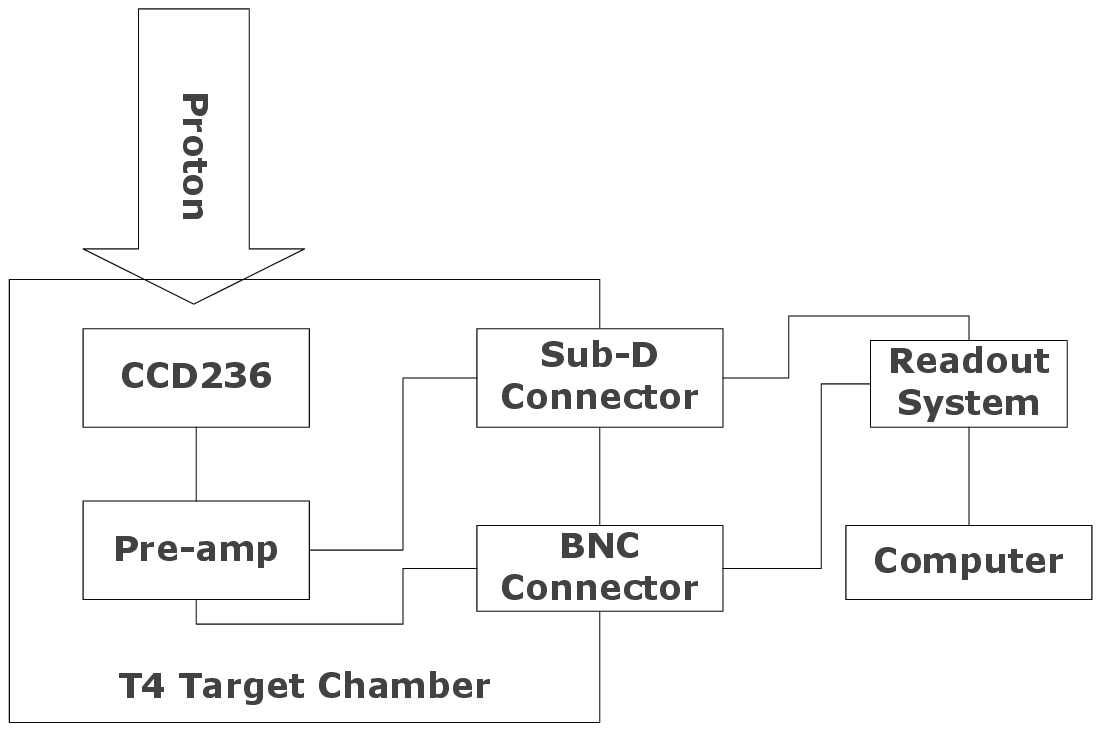}
\figcaption{\label{fig1} The schematic of the SCD proton irradiation experiment.}
\end{center}

\begin{center}
\includegraphics[origin=c,angle=90,width=8cm]{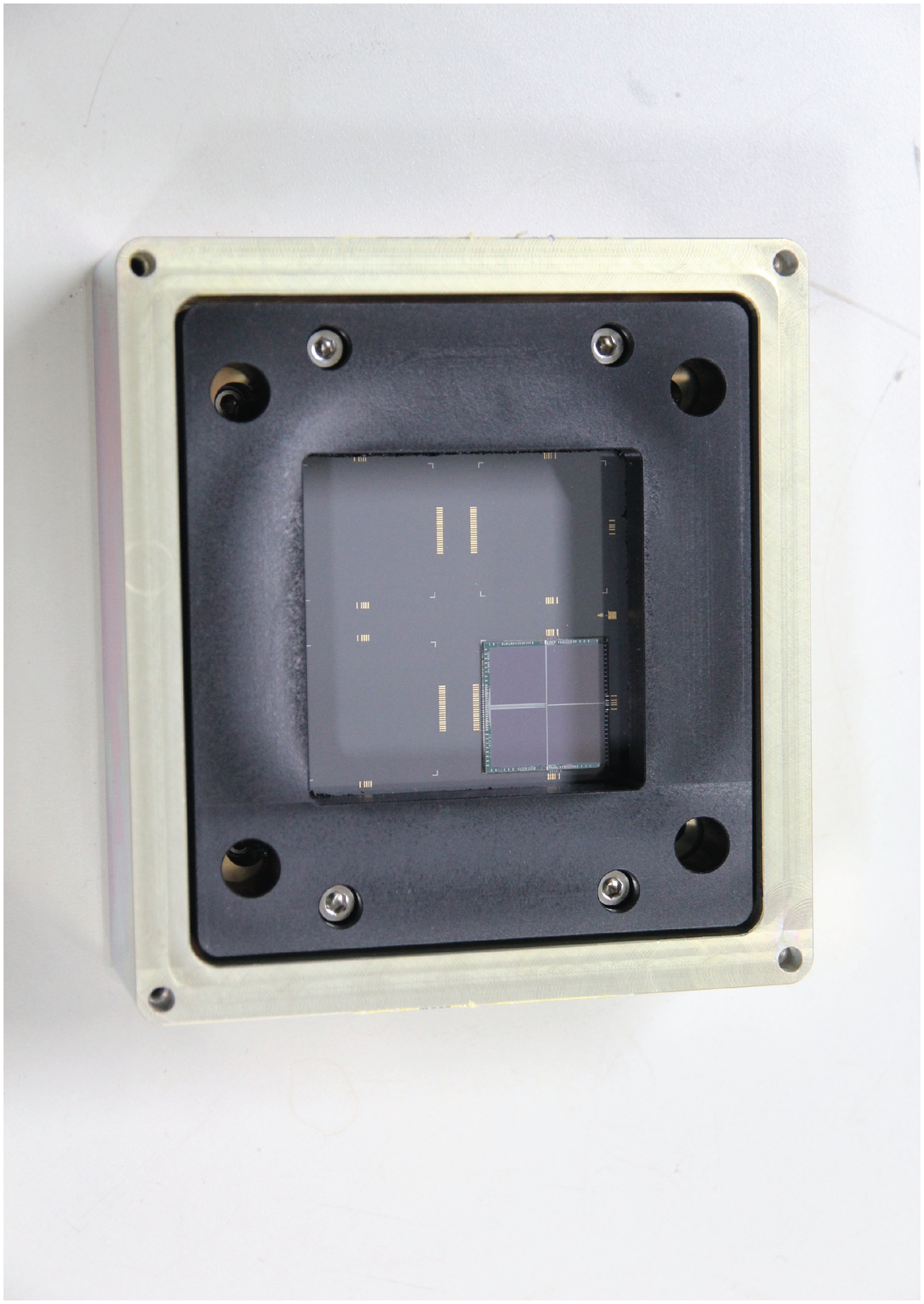}
\figcaption{\label{fig2} The SCD to be irradiated by protons.}
\end{center}

The SCD device was operated at 83.3 kHz. The operating voltages for the SCD are given in Table~\ref{tab1}.
\begin{center}
\tabcaption{\label{tab1} The operating voltages of the SCD.}
\footnotesize
\begin{tabular}{l|c}
\bottomrule
Parameter & Voltage/V \\
\hline
Substrate & 9.0 \\
Output Gate & 2.5 \\
Reset Gate2 & 17.0 \\
Output Drain & 30.0 \\
Clock Phase & 7.0 \\
\bottomrule
\end{tabular}
\end{center}

The SCD was irradiated by the protons at room temperature. The beam energy was measured to be 10 MeV at the SCD labeled as CCD236-20-4-D94.006. The proton fluence delivered to the SCD was $3\times10^{8}\mathrm{p}/\mathrm{cm}^{2}$ over 2 hours.

\section{Results and discussions}

As there were several difficulties in cooling down the SCD in the beam chamber, the SCD was tested in the vacuum tank at the Institute of High Energy Physics (IHEP). The SCD was glued at a copper plate cooled by liquid nitrogen as shown in Fig.~\ref{fig2}. The plate started to cool down after the chamber was vacuumized. When the SCD reached about $-110\,^{\circ}\mathrm{C}$, the cooling was stopped and the SCD started to warm slowly at the rate of less than $0.2\,^{\circ}\mathrm{C/min}$. The parameters, including the FWHM and the readout noise, were obtained by processing the spectra every 2 minutes continuously. A 1mCi $^{55}\mathrm{Fe}$ was also used to determine the performance in the chamber.\cite{lab8}
\begin{center}
\includegraphics[origin=c,angle=90,width=6cm]{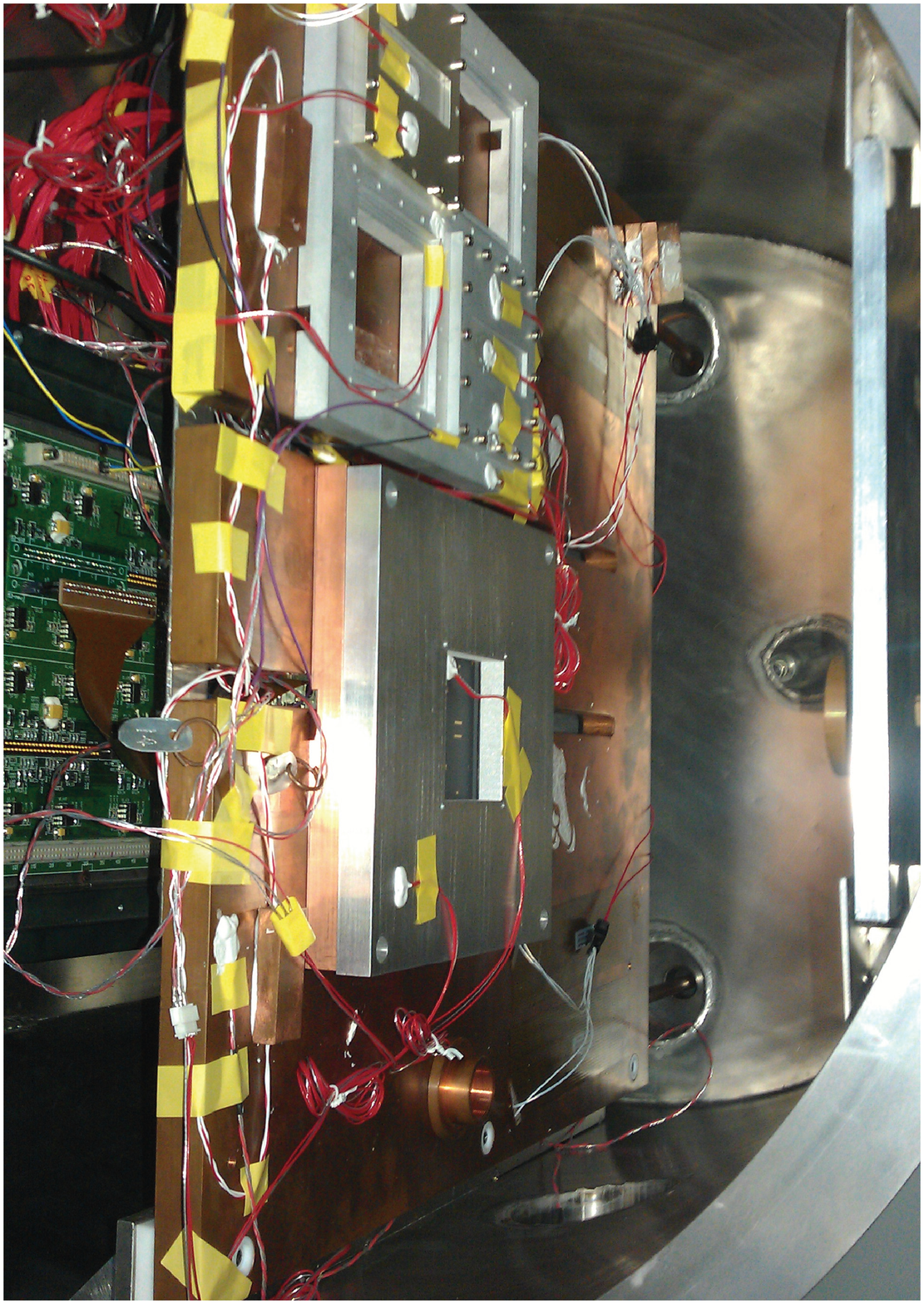}
\figcaption{\label{fig3} The vacuum chamber for SCD tests.}
\end{center}

 Two kinds of events are acquired from the SCD, the single events and the split events. The single events, which are called the single pixel events too, are the X-ray events that occupy a single pixel, while the split events are the X-ray events that occupy more than one pixel\cite{lab7}. In the data, the events are selected as follows. If there are events which amplitude are higher than threshold during more than 2 sequential drive clock, the events are split. And the rest of the events are single. Split events can be either abandoned or combined to reduce the background and improve the energy resolution.

 Three $^{55}\mathrm{Fe}$ spectra using the same data of the SCD are shown in Fig.~\ref{fig4}. Every event reading from the data shows directly without any processing in the raw spectrum. There are only single events in the single spectrum, whereas there are single events and split events which amplitudes are combined in the combined spectrum. Several peaks can be obtained from the spectra. From the plot, it can be learnt that both the single and the combined spectra can reduce the level of the background. However, as we know, each event of the split events is read out with one readout noise. Combining the split events will inevitably accumulate the readout noise at least twice, which causes worse energy resolution. Therefore, the single spectrum will be used for discussing the performance both before and after the irradiation.
\begin{center}
\includegraphics[width=8cm]{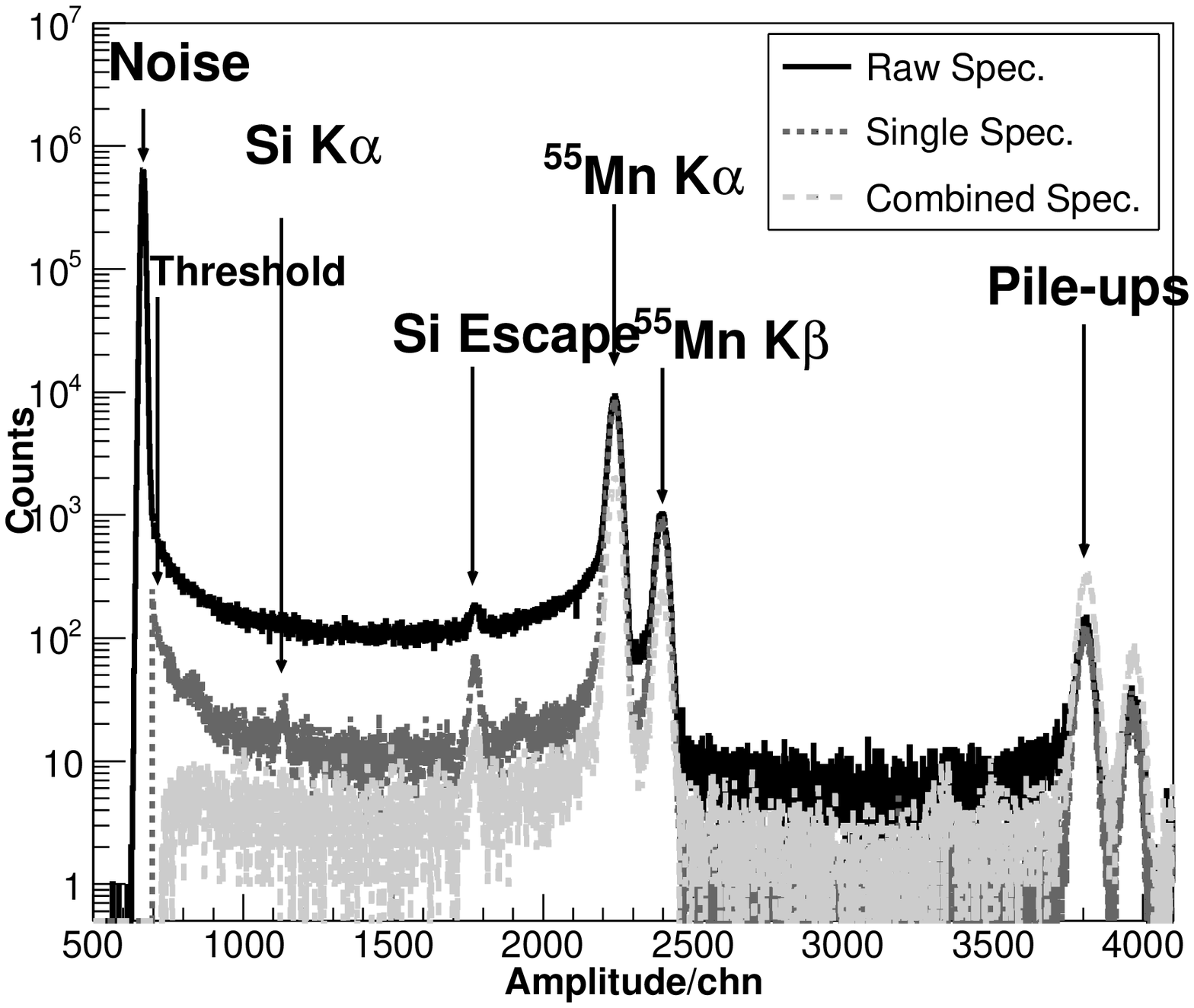}
\figcaption{\label{fig4} The $^{55}\mathrm{Fe}$ spectra of the SCD at $-75.1\,^{\circ}\mathrm{C}$.}
\end{center}

The spectra both before and after irradiation are shown simultaneously in Fig.~\ref{fig5} and Fig.~\ref{fig6}. It is obvious that the performance of the post-irradiation has been reduced a little at $-75.1\,^{\circ}\mathrm{C}$ in Fig.~\ref{fig5}, for the dark current has little impact on the resolution. When the operating temperature of the SCD is much higher, the post-irradiation spectrum is much worse as shown in Fig.~\ref{fig6}.
\begin{center}
\includegraphics[width=8cm]{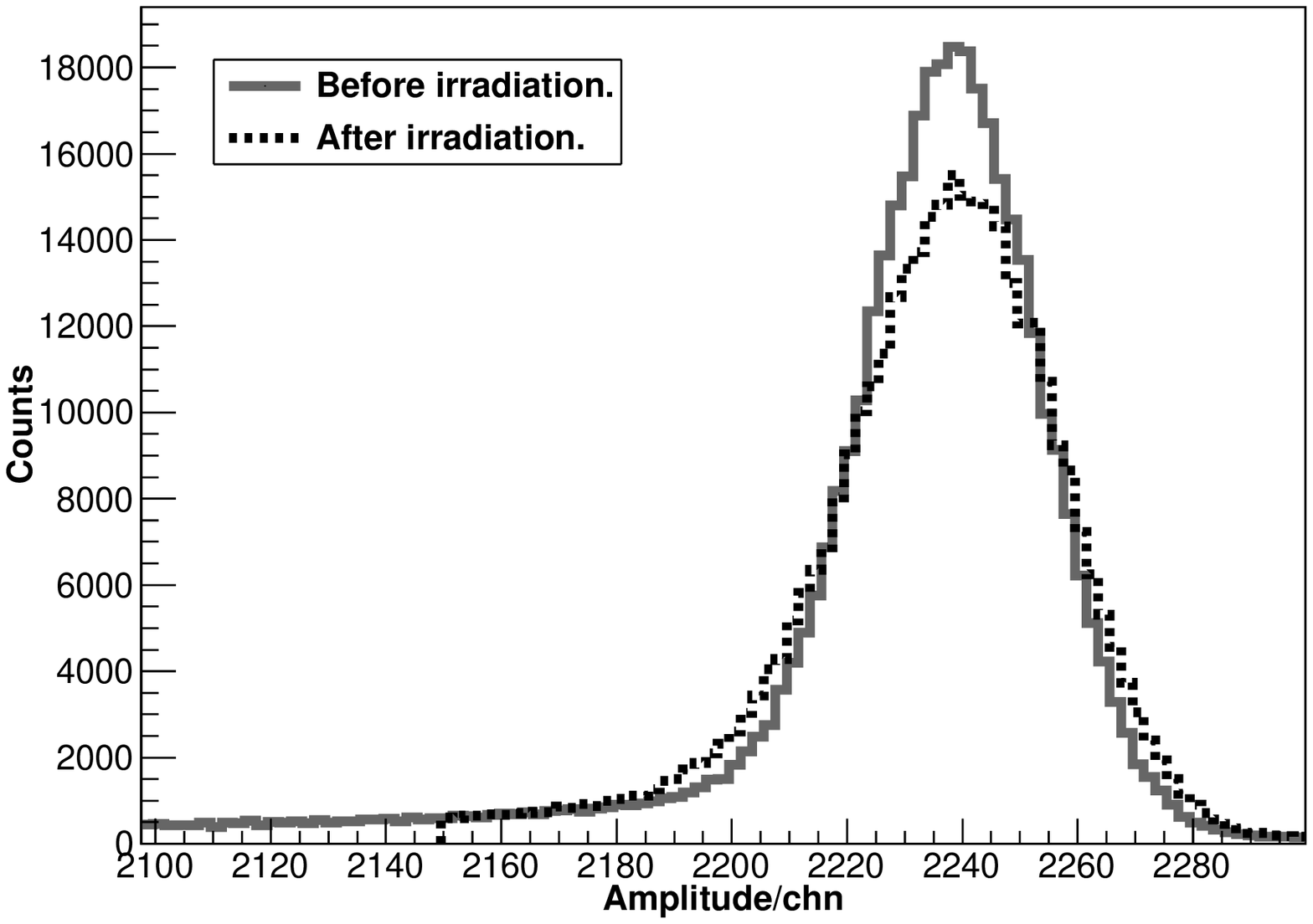}
\figcaption{\label{fig5} The $^{55}\mathrm{Mn}$ K$\alpha$ peak of the single events @$-75.1\,^{\circ}\mathrm{C}$.}
\end{center}
\begin{center}
\includegraphics[width=8cm]{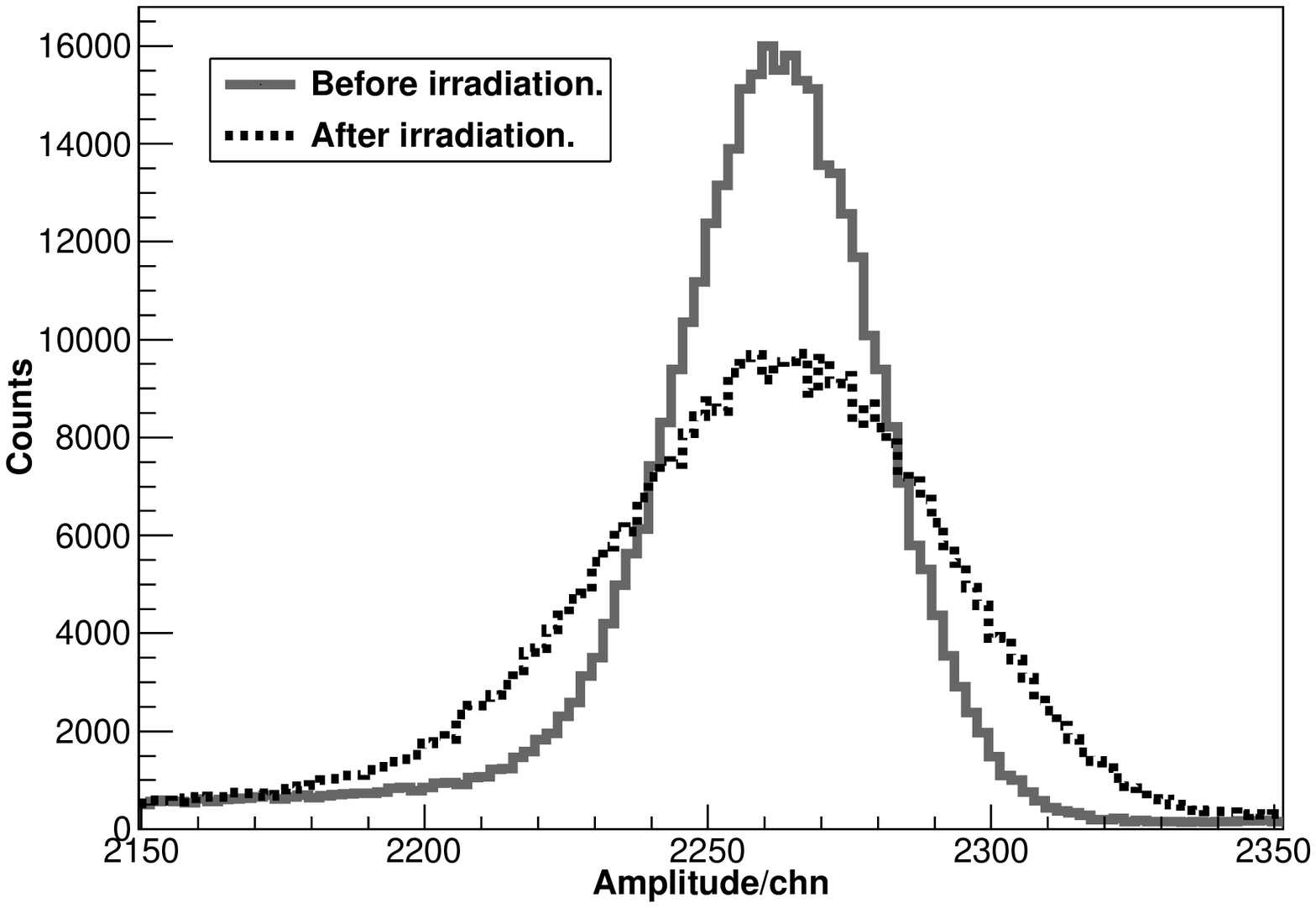}
\figcaption{\label{fig6} The $^{55}\mathrm{Mn}$ K$\alpha$ peak of the single events @$-31.1\,^{\circ}\mathrm{C}$.}
\end{center}

\subsection{Readout noise performance}

When there is no photon incoming, the output amplitude of the CCD is not zero, because there are still the dark current and other noises of the electronics in the output. The dark current noise depends on the detector itself. And the electronics noise relies on the electronics system. As the electronics system is placed at normal pressure and temperature in these two tests, the electronics noise can be considered constant. In Fig.~\ref{fig7}, when the temperature of the SCD is below $-60\,^{\circ}\mathrm{C}$, the value stands for the electronics noise. When the temperature of the SCD is above $-60\,^{\circ}\mathrm{C}$, the dark current becomes more and more higher as the temperature rises. Furthermore, the noise after the irradiation increases much faster than before the irradiation.
\begin{center}
\includegraphics[width=8cm]{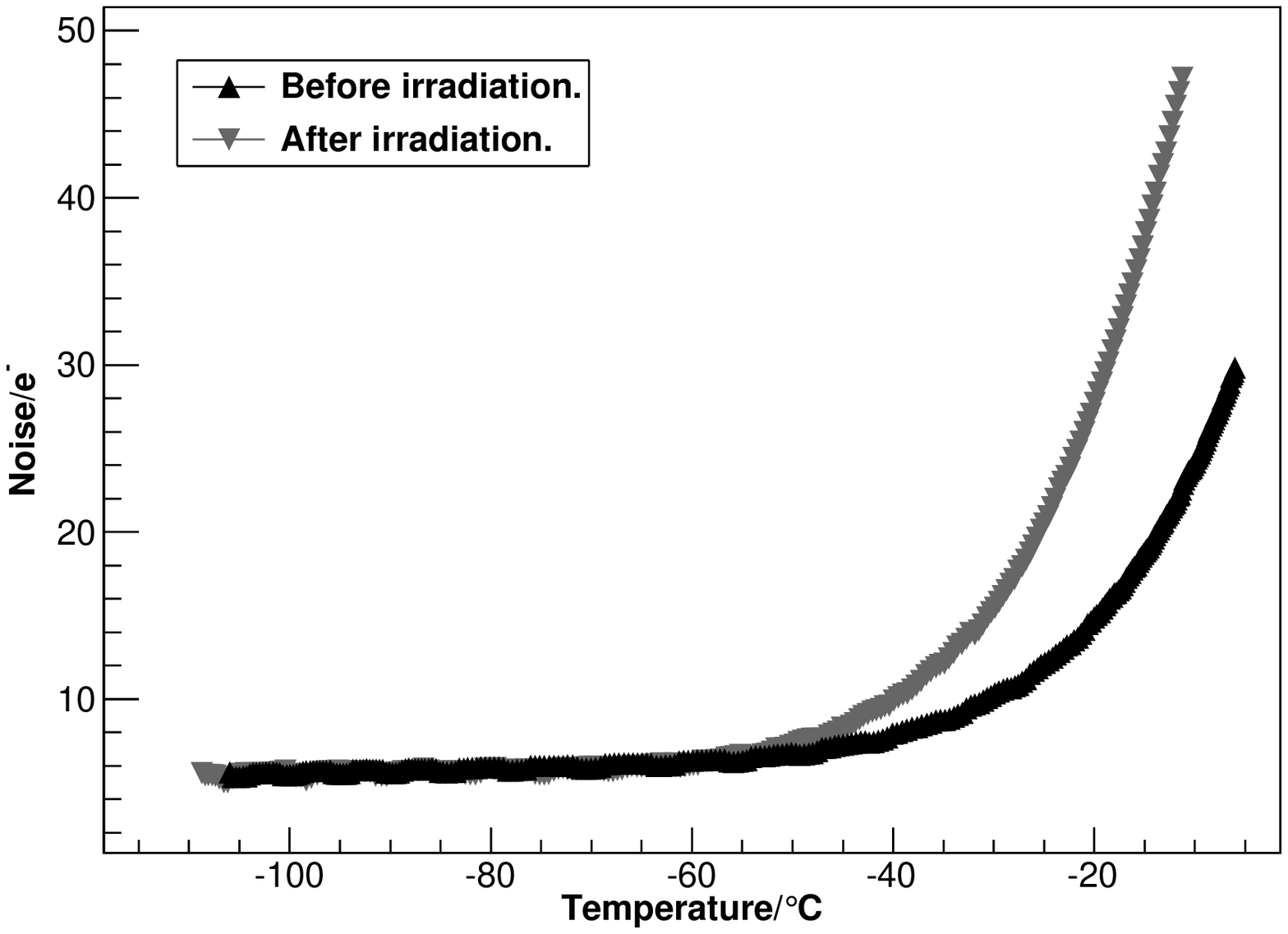}
\figcaption{\label{fig7} The readout noise before and after the irradiation vs. the SCD operating temperature.}
\end{center}

\subsection{Energy resolution performance}

The energy resolution of the FWHM at Mn K$\alpha$(5.9keV) is used to evaluate the effect of the proton irradiation. Through comparing the performance both before and after the irradiation, the FWHM after the irradiation is much higher than before, which is shown in Fig.~\ref{fig8}. The FWHM will be higher than 450 eV when the operating temperature of the SCD is $0\,^{\circ}\mathrm{C}$. However, it is unfortunate that the performance above $0\,^{\circ}\mathrm{C}$ can not be measured, as the amplitude of the signal is out of range. So it is a good choice to operate the SCD on orbit under $-20\,^{\circ}\mathrm{C}$.
\begin{center}
\includegraphics[width=8cm]{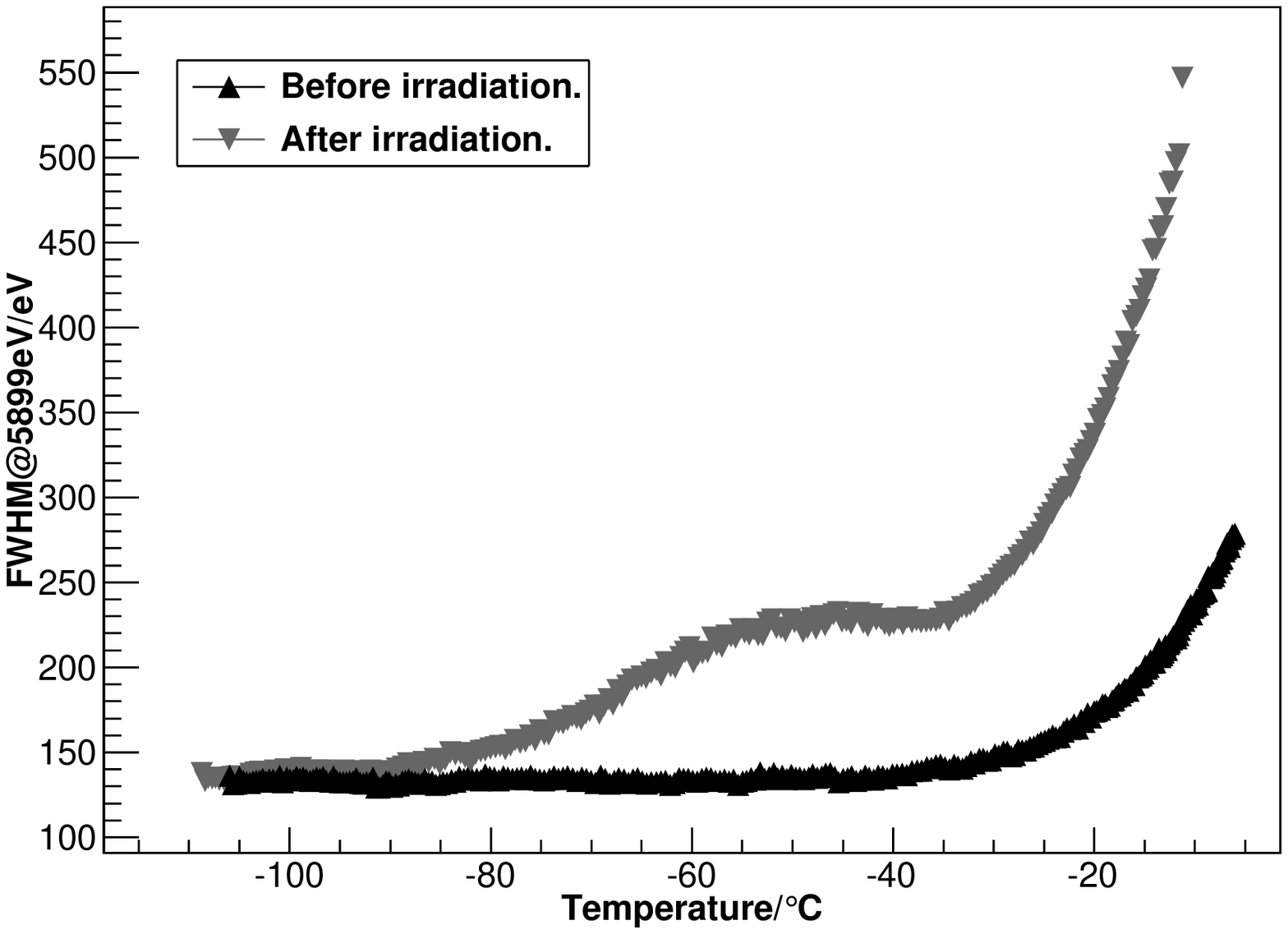}
\figcaption{\label{fig8} The FWHM before and after the irradiation vs. the SCD operating temperature.}
\end{center}

The energy resolution contains about three factors, including the shot noise, the CTI and the readout noise, which can be described as follows.

\begin{eqnarray}
\label{eq1}
FWHM &=& 2.355 \times{\omega}\times{\sqrt{\frac{F\times{E}}{\omega}+n_\mathrm{CTI}^{2}+n_\mathrm{noise}^{2}}}\nonumber\\[1mm]
\end{eqnarray}

In the equation, $\omega$ is the binding energy of the electrons and the hole, and it is 3.70 eV/e$^{-}$ in the discussion\cite{lab9}. $F$ stands for the fano factor, which is 0.12 here. $E$ is the energy of the incoming photons, which is 5899 eV for the Mn K$\alpha$. $n_\mathrm{CTI}$ is the noise caused by the CTI, $n_\mathrm{noise}$ is the readout noise.

\begin{center}
\includegraphics[width=8cm]{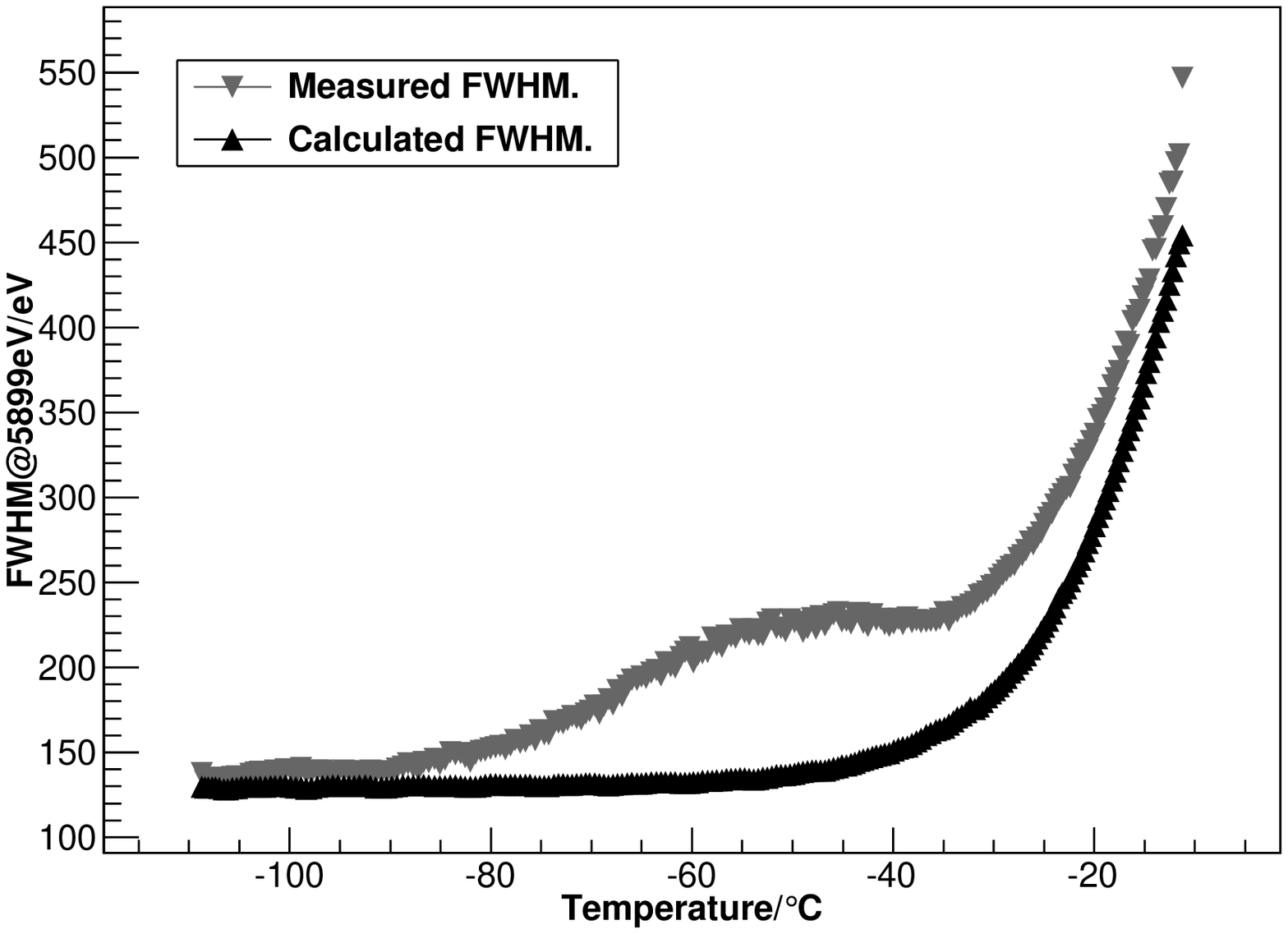}
\figcaption{\label{fig9} The measured and calculated FWHM ignoring CTI after the irradiation vs. the SCD operating temperature.}
\end{center}

\begin{center}
\includegraphics[width=8cm]{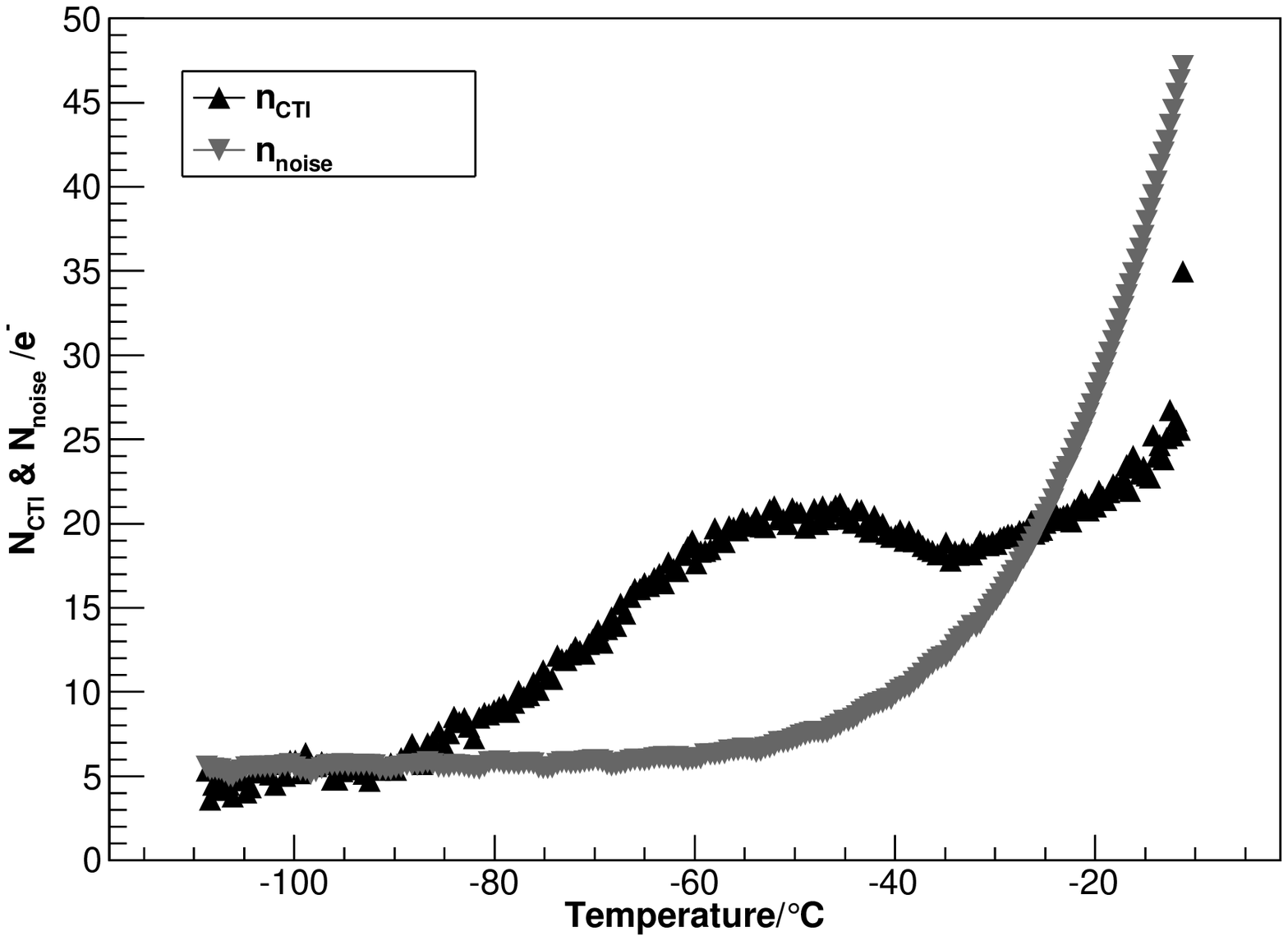}
\figcaption{\label{fig10} The $n_\mathrm{CTI}$ and $n_\mathrm{noise}$ after the irradiation vs. the SCD operating temperature.}
\end{center}

In Fig.~\ref{fig9}, the calculated FWHM curve is plotted without considering $n_\mathrm{CTI}$. And $n_\mathrm{noise}$ is from the curve after the irradiation in Fig.~\ref{fig6}. The curve can be split into 3 parts, which is directly illustrated in Fig.~\ref{fig10}. The first part is above $-30\,^{\circ}\mathrm{C}$, when the dark current dominates the energy resolution. The second part is from $-30\,^{\circ}\mathrm{C}$ to $-95\,^{\circ}\mathrm{C}$, when the CTI dominates. The third part is below $-95\,^{\circ}\mathrm{C}$, when both the CTI and the dark current contributes little. When it is below $-95\,^{\circ}\mathrm{C}$, the calculated FWHM is close to the measured value, which both the CTI and the dark current contributes little to the FWHM. When it is between $-60\,^{\circ}\mathrm{C}$ and $-95\,^{\circ}\mathrm{C}$, the dark current is near zero, while CTI contributes the most to the FWHM \cite{lab10}. When it is between $-30\,^{\circ}\mathrm{C}$ and $-60\,^{\circ}\mathrm{C}$, the dark current, which is not near zero any more, is still smaller than the CTI, which contributes most to the FWHM. When it is above $-30\,^{\circ}\mathrm{C}$, the noise of the dark current is so high that the noise of the CTI can be ignored. However, it has to be pointed out that there may be some errors in the section of the $n_\mathrm{CTI}$ curve above $-60\,^{\circ}\mathrm{C}$, because the dark current can also fill the traps in the proton-irradiated SCD, which leads to the decline of the $n_\mathrm{CTI}$ in the curve. Another test for illustrating that problem has been carried out too\cite{lab10}.

\section{Conclusion}

It can be concluded that the proton irradiation affects the performance of the SCD. Due to the performance of the SCD, the dark current increases significantly when it is above $-60\,^{\circ}\mathrm{C}$. In details, the dark current can be ignored when the SCD is below $-60\,^{\circ}\mathrm{C}$. The dark current noise of the proton-irradiated SCD increases faster than not irradiated. When it is below $-30\,^{\circ}\mathrm{C}$, the noise of the CTI dominates the energy resolution. When it is below $-95\,^{\circ}\mathrm{C}$, both the CTI and the dark current contributes little to the energy resolution. At that point, the best way to eliminate or weaken the proton irradiation effects is to lower the operating temperature of the SCD. In the test, the best temperature range for operating the proton-irradiated SCD is $-110\,^{\circ}\mathrm{C}$ to $-100\,^{\circ}\mathrm{C}$. Better performance can also be reached by lowering the operating temperature of the SCD on orbit.
\\

\end{multicols}

\clearpage

\end{document}